\documentclass[sigconf]{acmart}
\usepackage{multirow}
\usepackage{algorithm}
\usepackage{algorithmic}
\usepackage{subfigure}
\usepackage{multicol}

\usepackage{amsmath,amsfonts,bm}

\def\eqref#1{equation~\ref{#1}}

\def\1{\bm{1}}

\newcommand{\beps}{\boldsymbol\epsilon}

\def\rvepsilon{{\mathbf{\epsilon}}}

\def\rmI{{\mathbf{I}}}

\def\vzero{{\bm{0}}}

\def\vx{{\bm{x}}}

\def\mI{{\bm{I}}}

\DeclareMathAlphabet{\mathsfit}{\encodingdefault}{\sfdefault}{m}{sl}
\SetMathAlphabet{\mathsfit}{bold}{\encodingdefault}{\sfdefault}{bx}{n}

\def\gL{{\mathcal{L}}}

\def\gN{{\mathcal{N}}}

\newcommand{\E}{\mathbb{E}}

\AtBeginDocument{%
  \providecommand\BibTeX{{%
    \normalfont B\kern-0.5em{\scshape i\kern-0.25em b}\kern-0.8em\TeX}}}

\setcopyright{acmcopyright}

\copyrightyear{2022} 
\acmYear{2022} 
\setcopyright{acmcopyright}\acmConference[MM '22]{Proceedings of the 30th ACM International Conference on Multimedia}{October 10--14, 2022}{Lisboa, Portugal}
\acmBooktitle{Proceedings of the 30th ACM International Conference on Multimedia (MM '22), October 10--14, 2022, Lisboa, Portugal}
\acmPrice{15.00}
\acmDOI{10.1145/3503161.3547855}
\acmISBN{978-1-4503-9203-7/22/10}

\acmSubmissionID{455}

\begin{document}

\title{ProDiff: Progressive Fast Diffusion Model for High-Quality Text-to-Speech}







\author{Rongjie Huang$^{*}$, Zhou Zhao$^{\dagger}$, Huadai Liu$^{*}$, Jinglin Liu, Chenye Cui, Yi Ren}
\email{{rongjiehuang,zhaozhou,22160146,jinglinliu,chenyecui,rayeren}@zju.edu.cn}
\affiliation{Zhejiang University \country{}}



\begin{abstract}
\footnote{$^*$ Equal contribution. $\dagger$ Corresponding author.} Denoising diffusion probabilistic models (DDPMs) have recently achieved leading performances in many generative tasks. However, the inherited iterative sampling process costs hinder their applications to text-to-speech deployment. Through the preliminary study on diffusion model parameterization, we find that previous gradient-based TTS models require hundreds or thousands of iterations to guarantee high sample quality, which poses a challenge for accelerating sampling. In this work, we propose ProDiff, on progressive fast diffusion model for high-quality text-to-speech. Unlike previous work estimating the gradient for data density, ProDiff parameterizes the denoising model by directly predicting clean data to avoid distinct quality degradation in accelerating sampling. To tackle the model convergence challenge with decreased diffusion iterations, ProDiff reduces the data variance in the target site via knowledge distillation. Specifically, the denoising model uses the generated mel-spectrogram from an N-step DDIM teacher as the training target and distills the behavior into a new model with N/2 steps. As such, it allows the TTS model to make sharp predictions and further reduces the sampling time by orders of magnitude. Our evaluation demonstrates that ProDiff needs only 2 iterations to synthesize high-fidelity mel-spectrograms, while it maintains sample quality and diversity competitive with state-of-the-art models using hundreds of steps. ProDiff enables a sampling speed of 24x faster than real-time on a single NVIDIA 2080Ti GPU, making diffusion models practically applicable to text-to-speech synthesis deployment for the first time. Our extensive ablation studies demonstrate that each design in ProDiff is effective, and we further show that ProDiff can be easily extended to the multi-speaker setting. \footnote{Audio samples are available at \url{https://ProDiff.github.io/.}}

\end{abstract}




\maketitle
\section{Introduction}

Text-to-speech (TTS)~\cite{shen2018natural,ren2020fastspeech,huang2022generspeech} aims to generate almost human-like audios using text, which attracts broad interest in the machine learning community. Previous neural TTS models~\cite{wang2017tacotron,shen2018natural,li2019neural} first generate mel-spectrograms autoregressively from text and then synthesize speech from the generated mel-spectrograms using a separately trained vocoder~\cite{kong2020hifi,yamamoto2020parallel,you2021gan,huang2022sing}. They have been demonstrated to generate high-fidelity audio samples yet suffer from expensive computational costs. In recent years, non-autoregressive approaches~\cite{kim2020glow,popov2021grad,cui2021emovie} are proposed to generate speech audios with satisfactory speed. However, these models are criticized for other problems, e.g., the limited sample quality~\cite{ren2022revisiting} or sample diversity~\cite{ren2021portaspeech}.

In text-to-speech synthesis, our goal is mainly three-fold:
\begin{itemize}
    \item High-quality: to improve the naturalness of synthesized speech, the model should capture the details (frequency bins between two adjacent harmonics, unvoiced frames, and high-frequency parts) in natural speech.
    \item Fast: high generation speed is essential when considering real-time speech synthesis. This poses a challenge for all high-quality neural synthesizers.
    \item Diverse: to prevent the synthesized speech from being too dull and tedious when generating long speech, the model should be able to reduce mode collapse and avoid unimodal predictions.
\end{itemize}

As a blossoming class of generative models, denoising diffusion probabilistic models (DDPMs)~\cite{ho2020denoising,lam2021bilateral} have emerged to prove the capability to achieve leading performances in both image and audio synthesis~\cite{huang2022fastdiff,dhariwal2021diffusion}. However, the current development of DDPMs in speech synthesis is hampered by two major challenges:

\begin{itemize}
    \item While DDPMs inherently are gradient-based models with score matching objectives, a guarantee of high sample quality typically comes at the cost of hundreds to thousands of denoising steps. This prevents models from real-world deployment.
    \item When reducing refinement iterations, diffusion models show distinct degradation in model convergence due to the complex data distribution, leading to the blurry and over-smooth predictions in mel-spectrograms.
\end{itemize}

In this work, we start with a preliminary study on diffusion parameterization for text-to-speech. We find that the previous dominant diffusion models, which generate samples via estimating the gradient of the data density (denoted as gradient-based parameterization), require hundreds or thousands of iterations to guarantee high perceptual quality. When reducing the sampling steps, an apparent degradation in quality due to perceivable background noise is observed. On the contrary, the approach to parameterizing the denoising model through directly predicting clean data with a neural network (denoted as generator-based parameterization) has demonstrated its advantages in accelerating sampling from a complex distribution.

Based on these preliminary studies, we design better diffusion models for text-to-speech synthesis. In this paper, we propose \textbf{ProDiff}, on \textbf{pro}gressive fast \textbf{diff}usion model for high-quality text-to-speech; 1) To avoid significant degradation of perceptual quality when reducing reverse iterations, ProDiff directly predicts clean data $\vx$ and frees from estimating the gradient for score matching; 2) To tackle the model convergence challenge with decreased diffusion iterations, ProDiff reduces the data variance in the target side via knowledge distillation. Specifically, the denoising model uses the generated mel-spectrogram from an N-step DDIM teacher as the training target and distills the behavior into a new model with N/2 steps. As such, it allows the TTS model to make sharp predictions and further accelerates sampling by orders of magnitude.

Experimental results demonstrate that ProDiff achieves outperformed sample quality and diversity. ProDiff enjoys an effective sampling process that needs only 2 iterations to synthesize high-fidelity mel-spectrograms, 24x faster than real-time on a single NVIDIA 2080Ti GPU without engineered kernels. To the best of our knowledge, ProDiff makes diffusion models applicable to interactive, real-world speech synthesis applications at a low computational cost for the first time. The main contributions of this work are summarized as follows:

\begin{itemize}
    \item We analyze and compare different diffusion parameterizations in text-to-speech synthesis. Compared to the traditional gradient-based DDPMs with score matching objectives, the models that directly predict clean data show advantages in accelerating sampling from a complex distribution.
    \item We propose ProDiff, on progressive fast diffusion model for high-quality text-to-speech. Unlike estimating the gradient of data density, ProDiff parameterizes the denoising model by directly predicting clean data. To tackle the model convergence challenge in accelerating refinement, ProDiff uses the generated mel-spectrograms with reduced variance as target and makes sharper predictions. ProDiff is distilled from the behavior of the N-step teacher into a new model with N/2 steps, further decreasing the sampling time by orders of magnitude. 
    \item Experimental results demonstrate ProDiff needs only 2 iterations to synthesize high-fidelity mel-spectrograms, while it maintains sample quality and diversity competitive with state-of-the-art models using hundreds of steps. It makes diffusion models practically applicable to text-to-speech deployment for the first time.
\end{itemize}

\section{Background on Diffusion Models}
In this section, we introduce the theory of diffusion probabilistic model~\cite{ho2020denoising,lam2021bilateral,song2020denoising,song2020score}. Diffusion and reverse processes are given by diffusion probabilistic models, which could be used for the denoising neural networks $\theta$ to learn data distribution.

With the predefined fixed noise schedule $\beta$ and diffusion step $t$, we compute the corresponding constants respective to diffusion and reverse process:
\begin{equation}
    \small
    \alpha_{t}=\prod_{i=1}^{t} \sqrt{1-\beta_{i}} \quad \sigma_t=\sqrt{1-\alpha_t^2}
\end{equation}

\textbf{Diffusion process}  Similar as previous work~\cite{ho2020denoising,lam2021bilateral,song2020denoising}, we define the data distribution as $q(\vx_{0})$. The diffusion process is defined by a fixed Markov chain from data $\vx_0$ to the latent variable $\vx_T$:
\begin{equation}
    \small
    \label{diffusion}
q(\vx_{1},\cdots,\vx_T|\vx_0) = \prod_{t=1}^T q(\vx_t|\vx_{t-1}),
\quad\ \ 
\end{equation}

For a small positive constant $\beta_t$, a small Gaussian noise is added from $\vx_{t}$ to the distribution of $\vx_{t-1}$ under the function of $q(\vx_t|\vx_{t-1})$.

The whole process gradually converts data $\vx_0$ to whitened latents $\vx_T$ according to the fixed noise schedule $\beta_1,\cdots,\beta_T$.
\begin{equation}
    \small
q(\vx_t|\vx_{t-1}) := \gN(\vx_t;\sqrt{1-\beta_t}\vx_{t-1},\beta_t \rmI)
\end{equation}

\textbf{Reverse process}  The reverse process aims to recover samples from Gaussian noises, which is a Markov chain from $\vx_T$ to $\vx_0$ parameterized by shared $\theta$:
\begin{equation}
    \small
p_{\theta}(\vx_0,\cdots,\vx_{T-1}|x_T)=\prod_{t=1}^T p_{\theta}(\vx_{t-1}|\vx_t),
\end{equation}

where each iteration eliminate the Gaussian noise added in the diffusion process:
\begin{equation}
    \small
    p_{\theta}(\vx_{t-1}|\vx_t) := \mathcal{N}(\vx_{t-1};\mu_{\theta}(\vx_t,t), \sigma_{t}^2 \rmI)
\end{equation}

It has been demonstrated that diffusion models~\cite{dhariwal2021diffusion,xiao2021tackling} can learn diverse data distribution in multiple domains, such as images and time series. However, the main issue with the proposed neural diffusion process is that it requires up to thousands of iterative steps to reconstruct the target distribution during reverse sampling. In this work, we offer a progressive fast conditional diffusion model to reduce reverse iterations and enjoy computational efficiency.

\section{Diffusion model parameterization} \label{parameterization}
In this section, we discuss how to parameterize the reverse denoising model $\theta$ in a way for which the implied prediction. We classify current diffusion parameterization into two classes: 1) the denoising model learns the gradient of the data log density and predicts samples in the $\beps$ space, which we denote as the \textbf{\emph{Gradient-based method}}. 2) the denoising model directly predicts clean data $\vx_0$ and optimizes the sample reconstruction error, which we denote as the \textbf{\emph{Generator-based method}}.

\subsection{Gradient-based method}
The Stein score function is the gradient of the data log-density $\log p(\vx)$ with respect to the data $\vx$. Given the Stein score function $s(\cdot) = \nabla_{\vx} \log p(\vx)$, one can draw samples $\tilde{\vx} \sim p(\vx)$ from the corresponding density via Langevin dynamics, which can be interpreted as stochastic gradient ascent in the data space:

\begin{equation}
    \tilde\vx_{t+1}=\tilde\vx_{t}+\frac{\eta}{2} s(\tilde\vx_{t})+\sqrt{\eta} z_{t},
\end{equation}
where $\eta > 0$ is the step size, $z_{t} \sim \mathcal{N}(\vzero,\rmI)$.

A line of works named score matching neural networks~\cite{song2020score,song2020improved} learn the Stein score function $s(\cdot)$, and use Langevin dynamics for inference. For any step $t$, the denoising score matching objective takes the form:

\begin{equation}
    \E_{x \sim p(x)} \E_{\tilde{\vx} \sim q(\tilde{\vx}|x)}\left[\left \lVert s_{\theta}(\tilde{\vx})-\nabla_{\tilde{\vx}} \log q(\tilde{\vx}|x)\right\rVert_2^2\right],
    \end{equation}
where we have $\nabla_{\tilde{\vx}} \log q(\tilde{\vx}|x) =  -\frac{\beps}{\sigma_t}$, proportional to Gaussian noise $\beps$.

Simultaneously, another line of works named denoising diffusion probabilistic models (DDPMs)~\cite{ho2020denoising,dhariwal2021diffusion,kong2020diffwave} choose to parameterize the denoising model $\theta$ through directly predicting $\beps$ with a neural network $\beps_\theta$. 

Most recently, ~\citet{ho2020denoising} observe that denoising diffusion probabilistic models~\cite{sohl2015deep,dhariwal2021diffusion,kong2020diffwave} and score matching neural networks~\citep{song2020score,song2020improved} are closely related, which we denote as the \textbf{\emph{Gradient-based method}}. In this case, the training loss is usually defined as mean squared error in the $\beps$ space, and efficient training is optimizing a random term of $t$ with stochastic gradient descent:
\begin{equation}
    \small
    \label{eq: loss1}
    \gL_{\theta}^{\text{Grad}} = \left\lVert \beps_\theta\left(\alpha_t\vx_{0}+\sqrt{1-\alpha_t^2}\beps\right)-\beps\right\rVert_2^2, \beps\sim\gN(\vzero, \rmI)
\end{equation}

In a gradient-based diffusion model, a guarantee of high sample quality typically comes at the cost of hundreds to thousands of denoising steps, and thus the huge computational costs hinder its application in real-world text-to-speech deployment.
\subsection{Generator-based method}
Different from the aforementioned gradient-based diffusion models that required hundreds of steps with small $\beta_t$ to estimate the gradient for data density, diffusion models~\citep{salimans2022progressive,liu2022diffgan} can be interpreted as parameterizing the denoising model by directly predicting the clean data, which we denote as the \textbf{\emph{Generator-based method}}.


It is well-known that $\vx_{t}$ has different levels of perturbation, and hence using a single (gradient-based parameterization) network to predict $\vx_{t-1}$ directly at different $t$ may be difficult. In contrast, the generator-based diffusion model is free from estimating the gradient for data density. It only needs to predict unperturbed $\vx_{0}$ and then add back perturbation using the posterior distribution $q(\vx_{t-1}|\vx_t,\vx_0)$, which has been given in Appendix~\ref{Posterior}.

Specifically, in the generator-based diffusion models, $p_{\theta}(\vx_{0} |\vx_{t})$ is the implicit distribution imposed by the neural network $f_{\theta}(\vx_{t}, t)$ that outputs $\vx_{0}$ given $\vx_{t}$. And then $\vx_{t-1}$ is sampled using the posterior distribution $q(\vx_{t-1}|\vx_t,\vx_0)$ given $\vx_t$ and the predicted $\vx_0$. 

In this case, the training loss is defined as mean squared error in the data $\vx$ space, and efficient training is optimizing a random term of $t$ with stochastic gradient descent: 
\begin{equation}
    \small
    \label{eq: loss2}
    \gL_{\theta}^{\text{Gen}} = \left\lVert \vx_\theta\left(\alpha_t\vx_{0}+\sqrt{1-\alpha_t^2}\beps\right)-\vx_{0}\right\rVert_2^2, \beps\sim\gN(\vzero, \mI)
\end{equation}

Some recent works~\citep{salimans2022progressive,liu2022diffgan} choose to parameterize the denoising model through directly predicting $\vx_{0}$ with a neural network $f_\theta$. These generator-based methods advantage in accelerating sampling from a complex distribution.

\begin{figure*}[ht]
\centering
\includegraphics[width=0.9\textwidth]{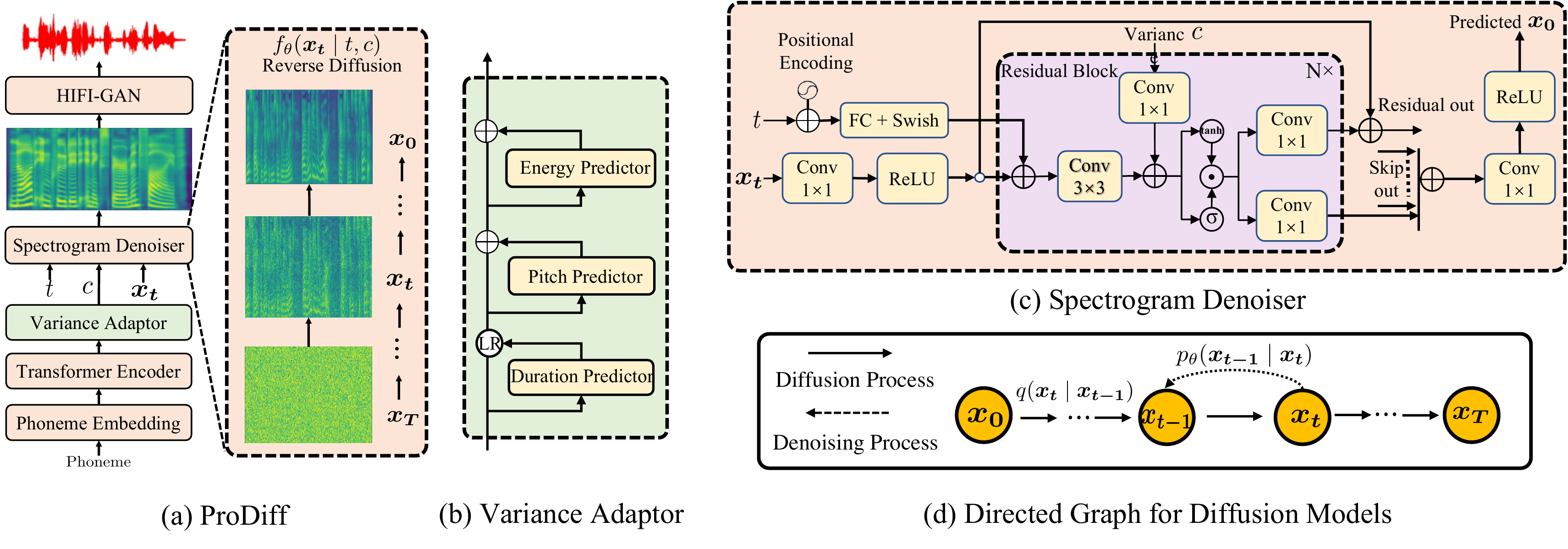}
\caption{The overall architecture for ProDiff. In subfigure (b), "LR" denotes the length regulator proposed in FastSpeech. In subfigure (c), The spectrogram denoiser $\theta$ takes noisy spectrogram $\vx_t$ as input and computes $f_{\theta}(\vx_t|t,c)$ conditioned on diffusion time index $t$ and variance $c$ (the output of the variance adaptor). We use the sinusoidal-like symbol, FC, Swish, LReLU and, $\bullet$ to denote the positional encoding,  fully-connected layer, swish activation function~\cite{ramachandran2017searching}, leaky rectified linear unit, and element-wise multiple operation. N is the number of residual layers.}
\label{fig:arch}
\end{figure*}
\section{ProDiff}
This section presents our proposed ProDiff, on progressive fast diffusion model for high-quality text-to-speech. We first describe the motivation of each design in ProDiff. Secondly, we introduce how to select a diffusion teacher model and distill from it; Then we describe the model 
architecture and training loss in ProDiff, following with the illustration of training and inference algorithms. 

\subsection{Motivation}
As a blossoming class of generative models, denoising diffusion probabilistic models (DDPMs) have emerged to prove their capability to achieve leading performances in both image and audio synthesis~\cite{dhariwal2021diffusion,san2021noise,kong2020diffwave,chen2020wavegrad}, while several challenges remain for industrial deployment: 
1) The most dominant diffusion TTS models estimate the gradient for data density with score matching objectives, where hundreds of iterations are required to guarantee high-quality synthesis. This prevents models from real-world deployment. 2) When reducing refinement iterations, diffusion models show distinct degradation in model convergence due to the complex data distribution. As such, the original denoising model cannot generate deterministic values, leading to blurry and over-smooth predictions in mel-spectrograms.

In ProDiff, we propose two key techniques to complement the above issues: 1) Utilizing the generator-based parameterization to speed up sampling. Through the preliminary comparisons of diffusion parameterization in Section~\ref{parameterization}, we conclude that the denoising model that predicts clean data has advantages in accelerating sampling from a complex distribution. 2) Reducing data variance in the target side via knowledge distillation. The denoising model uses the generated mel-spectrogram from an N-step DDIM teacher as the training target and distills the behavior into a new model with N/2 steps. As such, it allows the TTS model to make a sharp prediction and further reduces the sampling time by orders of magnitude.

To conclude, ProDiff maintains the sample quality and frees diffusion models from hundreds of iterative refinements, which is for the first time applicable to interactive, real-world applications. 

\subsection{Select a teacher} \label{select_teacher}
In this subsection, we describe how to select an expected teacher for knowledge distillation. The teacher model is supposed to achieve the fast, high-quality, and diverse text-to-speech synthesis, and thus the distilled student could inherit its powerful capability. Through the preliminary analyses (see Section~\ref{parameterization}) and experiments (see Section~\ref{exp_parameterization}) on diffusion parameterization, we empirically find that the 4-step generator-based diffusion model strikes a proper balance between perceptual quality and sampling speed. As such, we pick up a generator-based diffusion model $\theta$ with 4 diffusion steps as the teacher.

\subsection{Distill from teacher}
Denoising diffusion implicit model (DDIM)~\citep{song2020denoising} formulates a non-Markovian generative process that accelerates the inference while keeping the same training procedure as denoising diffusion probabilistic model (DDPM). Inspired by ~\citet{salimans2022progressive}, we utility the sampler to directly predict the coarse-grained mel-spectrogram with reduced variance in the diffusion process. 

Specifically, we first initialize ProDiff with a copy of the teacher mentioned in Section~\ref{select_teacher}, using both the same parameters and model definition. We sample data from the training set and add noise to it as original. Differently, we get the target value $\hat \vx_{0}$ for the denoising model by running 2 DDIM sampling steps using the teacher instead of the original data $\vx_{0}$. Further, we halve the required steps by making a single DDIM step of the student match 2 DDIM steps of the teacher. As illustrated in Algorithm~\ref{alg: training2}, the implementation of training the student model with knowledge distillation stays very close to the original Algorithm~\ref{alg: training_generator} in the appendix.


\subsection{Architecture}
We build the basic architecture upon FastSpeech 2~\cite{ren2020fastspeech}, which is one of the most popular models in non-autoregressive TTS. As illustrated in Figure~\ref{fig:arch}, ProDiff consists of a phoneme encoder, variance adaptor, and spectrogram denoiser. The phoneme encoder converts a sequence of phoneme embedding into a hidden sequence. Then, the variance adaptor predicts the duration of each phoneme to regulate the length of the hidden sequence into the length of speech frames. Furthermore, the variance adaptor indicates different variances in speech, such as pitch and energy. Finally, the spectrogram denoiser iteratively refines the length-regulated hidden sequence into mel-spectrograms. More details have been attached in Appendix~\ref{appendix:arch}.

\textbf{Encoder and Variance Adaptor}
The phoneme encoder is composed of feed-forward transformer blocks (FFT blocks) based on the transformer architecture~\cite{vaswani2017attention}. The encoder comprises a pre-net, transformer blocks with multi-head self-attention, and the final linear projection layer. In variance adaptor, the duration, pitch, and energy predictors share a similar model structure consisting of a 2-layer 1D-convolutional network with ReLU activation, each followed by the layer normalization and the dropout layer, and an extra linear layer to project the hidden states into the output sequence.

\textbf{Spectrogram Denoiser}
Following~\citet{liu2021diffsinger}, we adopt a non-causal WaveNet~\cite{oord2016wavenet} architecture to be our spectrogram denoiser. The decoder comprises a 1x1 convolution layer and $N$ convolution blocks with residual connections to project the input hidden sequence with 256 channels. For any step $t$, we use the cosine schedule $\beta_{t}=\cos (0.5 \pi t)$. 

\subsection{Training Loss}
Several objectives have been used to optimize the ProDiff model.

\textbf{Sample Reconstruction Loss}
Instead of using the original clean data $\vx_{0}$ in training ProDiff, we get the target value $\hat \vx_{0}$ with reduced variance by running 2 DDIM sampling steps of the teacher: 

\begin{equation}
  \small
  \label{loss_reconstruct}
    \gL_{\theta} = \left\lVert \vx_\theta\left(\alpha_t\vx_{0}+\sqrt{1-\alpha_t^2}\beps\right)-\hat\vx_{0}\right\rVert_2^2, \beps\sim\gN(\vzero, \rmI)
\end{equation}

\textbf{Structural Similarity Index (SSIM) Loss}
Structural Similarity Index (SSIM)~\cite{wang2004image} is one of the state-of-the-art perceptual metrics to measure image quality, which can capture structural information and texture. The value of SSIM is between 0 and 1, where 1 indicates perfect perceptual quality relative to the ground truth. Inspired by~\cite{ren2022revisiting}, we adopt structural similarity index (SSIM) loss in training TTS models:

\begin{equation}
  \small
  \label{loss_ssim}
    \gL_{\operatorname{SSIM}} = 1-\operatorname{SSIM}\left(\vx_\theta\left(\alpha_t\vx_{0}+\sqrt{1-\alpha_t^2}\beps\right), \hat\vx_{0}\right)
\end{equation}

\textbf{Variance Reconstruction Loss}
To promote the naturalness and expressiveness in generated speech, we provide more acoustic variance information including pitch, duration and energy. Variance reconstruction losses are also added to train the acoustic generator. 

\begin{equation}
  \small
  \label{loss_variance}
\gL_{p} = \lVert p - \hat p \rVert_2^2, \gL_{e} = \lVert e - \hat e \rVert_2^2, \gL_{dur} = \lVert d - \hat d \rVert_2^2,
\end{equation}
where we use $d$, $e$ and $p$ to denote the target duration, energy and pitch, respectively, and use $\hat d$, $\hat e$ and $\hat p$ to denote the corresponding predicted values. The loss weights are all set to be 0.1.

\subsection{Training and Inference Procedures}
The training and sampling algorithm of ProDiff have been illustrated in Algorithm~\ref{alg: training2} and Algorithm~\ref{alg: sampling}, respectively.

\paragraph{\textbf{Training}}
\par The final loss term in training ProDiff consist of the following parts: 1) sample reconstruction loss $\gL_{\theta}$: MSE between the predicted and the target mel-spectrogram according to Eq.~\ref{loss_reconstruct}; 2) structural similarity index (SSIM) loss $\gL_{\operatorname{SSIM}}$: one minus the SSIM index between the predicted and the target mel-spectrogram according to Eq.~\ref{loss_ssim}. 3) variance reconstruction loss, $\gL_{dur}, \gL_{p}, \gL_{e}$: MSE between the predicted and the target phoneme-level duration, pitch spectrogram, and energy value according to Eq.~\ref{loss_variance};

\paragraph{\textbf{Inference}}
\par  In inference, ProDiff iteratively predicts unperturbed $\vx_{0}$ and then adds back perturbation using the posterior distribution, and thus finally it generates high-fidelity mel-spectrograms with increasing details. Specifically, the denoising model $f_{\theta}(\vx_t|t,c)$ first predicts $\hat \vx_{0}$, and then $\vx_{t-1}$ is sampled using the posterior distribution $q(\vx_{t-1}|\vx_t,\vx_0)$ given $\vx_{t}$ and the predicted $\hat \vx_{0}$. In final, the generated spectrogram $\vx_{0}$ is transformed to waveforms using a pre-trained vocoder. The posterior distribution has been given in Appendix~\ref{Posterior}.

\begin{algorithm}[ht]
    \centering
    \caption{Training ProDiff via knowledge distillation}\label{alg: training2}
    \begin{algorithmic}[1]
     \STATE \textbf{Require}: The ProDiff teacher $\zeta$ (diffusion steps $T_1$), ProDiff $\theta$ (diffusion steps $T_2$), and variance condition $c$.
    \REPEAT 
    \STATE Sample $\vx_{0} \sim q_{data}$, $\rvepsilon\sim\gN(\vzero,\mI)$
    \STATE $S = T_1 / T_2$, Sample $t\sim S \cdot \mathrm{Unif}(\{1,\cdots,T_2\})$
    \STATE $\vx_t = \alpha_{t} \vx_{0}+ \sigma_t \rvepsilon$
    \STATE $t^{\prime} = t-S/2, t^{\prime \prime} = t - S$
    \STATE $\vx_t^{\prime} = \alpha_{t}^{\prime} f_\zeta(\vx_t|c, t)+ \frac{\sigma_{t^{\prime}}}{\sigma_t} (\vx_t - \alpha_{t} f_\zeta(\vx_t|c, t))$
    \STATE $\vx_t^{\prime\prime} = \alpha_{t}^{\prime\prime} f_\zeta(\vx_t^\prime|c, t^\prime)+ \frac{\sigma_{t^{\prime\prime}}}{\sigma_t^\prime} (\vx_t^\prime - \alpha_{t} f_\zeta(\vx_t^\prime|c, t^\prime))$
    \STATE $\hat \vx_{0} = \frac{\vx_t^{\prime\prime} - (\sigma_{t^{\prime\prime}} / \sigma_t) \vx_t}{\alpha_t^{\prime\prime} - (\sigma_{t^{\prime\prime}} / \sigma_t) \alpha_t}$
    \STATE Take gradient descent steps on $\nabla_{\theta}(\gL_{\theta} + \gL_{\operatorname{SSIM}} + \gL_{p} + \gL_{e} + \gL_{dur})$. 
    \UNTIL{ProDiff converged}
    \end{algorithmic}
    \end{algorithm}

      \begin{algorithm}[ht]
        \centering
        \caption{Sampling}\label{alg: sampling}
        \begin{algorithmic}[1]
         \STATE \textbf{Input}: ProDiff $\theta$, and variance condition $c$.
        \STATE Sample $\vx_{T} \sim \gN(\vzero,\mI)$
        \FOR{$t=T,\cdots,1$}
        \STATE $\hat \vx_0 = f_{\theta}(\vx_t|t,c)$
        \STATE Sample $\vx_{t-1}\sim p_\theta(\vx_{t-1}|\vx_t)= q(\vx_{t-1}|\vx_t, \hat \vx_0)$
        \ENDFOR
        \RETURN $\vx_0$
        \end{algorithmic}
        \end{algorithm}

\section{Related works}

\subsection{Text-to-Speech}
Text-to-speech (TTS) models convert input text or phoneme sequence into mel-spectrogram (e.g., Tacotron~\cite{wang2017tacotron}, FastSpeech~\cite{ren2019fastspeech}), which is then transformed to waveform using a separately trained vocoder~\cite{kong2020hifi,huang2021multi}, or directly generate waveform from text (e.g., EATS~\cite{donahue2020end} and VITS~\cite{kim2021conditional}). Early autoregressive models~\cite{wang2017tacotron,li2019neural} sequential generate a sample and suffer from slow inference speed. Several works~\cite{ren2019fastspeech,kim2020glow} have been proposed to generate mel-spectrogram frames in parallel, which speed up mel-spectrogram generation over autoregressive TTS models, while preserving the quality of synthesized speech. Recently proposed Diff-TTS~\cite{jeong2021diff}, Grad-TTS~\cite{popov2021grad}, and DiffSpeech~\cite{liu2021diffsinger} inherently are gradient-based models with score matching objectives to generate high-quality samples, while the iterative sampling process costs hinder their applications to text-to-speech deployment. Unlike the dominant gradient-based TTS models mentioned above, ProDiff directly predicts clean data and frees from estimating the gradient for score matching. The proposed model avoids significant perceptual quality degradation when reducing reverse iterations, and thus it maintains sample quality competitive with state-of-the-art models using hundreds of steps. 

\subsection{Diffusion Probabilistic Models}
Denoising diffusion probabilistic models (DDPMs)~\cite{ho2020denoising,song2020denoising,lam2021bilateral} are likelihood-based generative models that have recently succeeded to advance the state-of-the-art results in several important domains including image synthesis~\cite{dhariwal2021diffusion,song2020denoising}, audio synthesis~\cite{huang2022fastdiff}, and 3D point cloud generation~\cite{luo2021diffusion}, and have proved its capability to produce high-quality samples. One major drawback of diffusion or score-based models is the slow sampling speed due to many iterative steps. To alleviate this issue, multiple methods have been proposed, including learning an adaptive noise schedule~\cite{lam2022bddm}, introducing non-Markovian diffusion processes~\cite{song2020denoising}, and using adversarial learning for reducing iterations~\cite{xiao2021tackling}. To conclude, all these methods focus on the image domain, while audio data is different for its long-term dependencies and strong condition. ~\citet{liu2022diffgan} proposes a denoising diffusion generative adversarial networks (GANs) to achieve high-fidelity and efficient text-to-spectrogram synthesis. Differently, our work focuses on designing progressive fast diffusion models for text-to-speech synthesis without unstable adversarial learning procedure, which has been relatively overlooked.

\subsection{Knowledge distillation}
Knowledge distillation has been demonstrated for its efficiency in simplifying the data distribution. In non-autoregressive machine translation~\cite{gu2017non}, sequence-level knowledge distillation~\cite{kim2016sequence} has achieved good performance in transferring the knowledge from the teacher model to the student. In non-autoregressive text-to-speech synthesis~\cite{ren2019fastspeech}, researchers alleviate the one-to-many mapping problem by using the generated mel-spectrogram from an autoregressive teacher model as the training target. Recently, ~\citet{salimans2022progressive} utilize knowledge distillation and iteratively halve the diffusion steps of the continuous diffusion model, accelerating iterative refinement to a large extent. In contrast, ProDiff with discrete schedules adopts knowledge distillation to reduce data variance and promote training convergence. 
\section{Experiments}

\subsection{Experimental Setup}
\textbf{Dataset} For a fair and reproducible comparison against other competing methods, we use the benchmark LJSpeech dataset~\cite{LJSpeech}. LJSpeech consists of 13,100 audio clips of 22050 Hz from a female speaker for about 24 hours in total. We convert the text sequence into the phoneme sequence with an open-source grapheme-to-phoneme conversion tool~\cite{sun2019token}~\footnote{\url{https://github.com/Kyubyong/g2p}}. Following the common practice~\citep{chen2021adaspeech,min2021meta}, we conduct preprocessing on the speech and text data: 1) extract the spectrogram with the FFT size of 1024, hop size of 256, and window size of 1024 samples; 2) convert it to a mel-spectrogram with 80 frequency bins; and 3) extract F0 (fundamental frequency) from the raw waveform using Parselmouth tool~\footnote{\url{https://github.com/YannickJadoul/Parselmouth}}.

\noindent \textbf{Model Configurations}
ProDiff consists of 4 feed-forward transformer blocks for the phoneme encoder. In the pitch encoder, the size of the lookup table and encoded pitch embedding are set to 300 and 256. The hidden channel is set to 256. In the denoiser, we set $N=20$ to stack 20 layers of convolution with the kernel size 3, and we set the dilated factor to 1 (without dilation) at each layer. We have attached more detailed information on the model configuration in Appendix~\ref{appendix:arch}.

\noindent \textbf{Training and Evaluation} We train the ProDiff teacher with $T_1=4$ diffusion steps and take the converged teacher to train ProDiff with $T_2=2$ diffusion steps via knowledge distillation. The diffusion probabilistic models have been trained for 200,000 steps using 1 NVIDIA 2080Ti GPU with a batch size of 64 sentences. The adam optimizer is used with $\beta_{1}=0.9, \beta_{2}=0.98, \epsilon=10^{-9}$. We utilize HiFi-GAN\cite{kong2020hifi} (V1) as the vocoder to synthesize waveform from the generated mel-spectrogram in all our experiments. 
To evaluate the perceptual quality, we conduct crowd-sourced human evaluations with MOS (mean opinion score) on the testing set via Amazon Mechanical Turk, which is rated from 1 to 5 and reported with the 95\% confidence intervals (CI). We further include objective evaluation metrics, such as MCD~\cite{kubichek1993mel}, STOI~\cite{taal2010short}, and PESQ~\cite{rix2001perceptual} to evaluate the compatibility between the spectra of two audio sequences. To evaluate the sampling speed, we implement real-time factor (RTF) assessment on a single NVIDIA 2080Ti GPU. In addition, we employ two metrics NDB and JS~\cite{richardson2018gans} to explore the diversity of generated mel-spectrograms. More information on evaluation has been attached in Appendix~\ref{appendix:evaluation}.

\begin{table}[ht]
    \centering
    \small
    \begin{tabular}{l|ccc|ccc}
        \toprule
   \bfseries \multirow{2}{*}{Method} & \multicolumn{3}{c|}{\bfseries Gradient-Based} & \multicolumn{3}{c}{\bfseries Generator-Based} \\
                       & \bfseries RTF($\downarrow$) &\bfseries MOS ($\uparrow$) & \bfseries JS($\downarrow$) & \bfseries RTF ($\downarrow$)& \bfseries MOS ($\uparrow$) & \bfseries JS($\downarrow$) \\
    \midrule
    GT                  &  /       &  4.39$\pm$0.07  & /  & /      & 4.39$\pm$0.07  & / \\
    GT (voc.)           &  /       &  4.22$\pm$0.05  & /  & /      & 4.22$\pm$0.05  & / \\
    \midrule
    128 iter            & 1.21     & 4.09$\pm$0.04 &0.014 & 1.19  & 4.09$\pm$0.05 &0.009\\
    64 iter             & 0.65     & 4.08$\pm$0.05 &0.018 & 0.60  & \multirow{5}{*}{4.07$\pm$0.05$\dagger$} &0.009  \\
    32 iter             & 0.34     & 4.06$\pm$0.05 &0.011 & 0.31  &  &0.012  \\
    16 iter             & 0.17     & 3.95$\pm$0.05 &0.012 & 0.18  &  &0.010  \\
    8 iter              & 0.10     & 3.63$\pm$0.07 &0.043 & 0.10  & &0.014\\
    4 iter              & 0.06     & 3.25$\pm$0.08 &0.099 & \bfseries 0.06  & &\bfseries 0.014  \\
    2 iter              & 0.04     & 2.96$\pm$0.08 &0.179 & 0.04  & 3.98$\pm$0.04 &0.023\\
    \bottomrule
    \end{tabular}
    \vspace{1mm}
    \caption{Comparison between gradient-based and generator-based diffusion models with varying diffusion steps. Note that the iterative refinement steps in training and inference remain the same. $\dagger$: Raters could hardly tell differences between these samples, and thus we provide the averaged result.}
    \label{table:mos1}
    \end{table}
    
\subsection{Preliminary Analyses on Diffusion Parameterization} \label{exp_parameterization}

We compare and examine both choices (i.e., gradient-based or generator-based diffusion parameterization) with varying diffusion steps in Section~\ref{parameterization}, and conduct evaluations in terms of quality, diversity, and sampling speed. The results have been shown in Table~\ref{table:mos1}, and we also visualize the mel-spectrograms generated in Appendix~\ref{appendix:visualizations}. We have some observations from the results: 1) With a large distribution of noise schedule, gradient-based or generator-based diffusion models could synthesize high-fidelity speech samples with similar results. 2) When reducing iterative steps ($T\leq 16$), a distinct degradation due to the perceivable background noise is observed in gradient-based TTS models. In contrast, the generator-based TTS models which parameterize the denoising model by directly predicting clean data, still maintain sample quality and avoid significant degradation, which is expected for the following reasons: 
\begin{itemize}
    \item It is well-known that noisy data $\vx_{t}$ has different levels of perturbation in diffusion models, and hence using a neural network to predict $\vx_{t-1}$ directly at different $t$ may be difficult. As such, the gradient-based model requires hundreds of diffusion steps with small $\beta_t$ to estimate the gradient of the data density, and the distinct degradation is observed when accelerating refinement iterations.
    \item In contrast, the generator-based diffusion model is free from estimating the gradient for data density, which only needs to predict unperturbed $\vx_{0}$ and then add back perturbation using the posterior distribution $q(\vx_{t-1}|\vx_t,\vx_0)$. Therefore, the generator-based diffusion parameterization advantages in avoiding significant perceptual quality degradation when accelerating sampling from a complex distribution. 
\end{itemize}

\begin{table*}[ht]
\centering
  \begin{tabular}{l|cccc|cc|c}
  \toprule
  \multirow{2}{*}{\bfseries Method} & \multicolumn{4}{c|}{\bfseries Audio Quality} & \multicolumn{2}{c|}{\bfseries Sample Diversity}  & \multicolumn{1}{c}{\bfseries Speed} \\
                          & \bfseries MOS ($\uparrow$) & \bfseries MCD ($\downarrow$) & \bfseries STOI ($\uparrow$) & \bfseries PESQ ($\uparrow$) &\bfseries NDB($\downarrow$)   & \bfseries JS($\downarrow$) & \bfseries RTF($\downarrow$)   \\
  \midrule
  GT                                              & 4.41$\pm$0.06      &  / &  /  &  /& / &  /  & /  \\
  GT(voc.)                                        & 4.25$\pm$0.06      & 1.08  & 0.95 & 3.18    & 0.23  &0.002  & / \\
  \midrule            
  Tacotron 2~\cite{shen2018natural}               & 3.90$\pm$0.07   & 5.30  & 0.18 & 1.14 & 0.88 & 0.022  &0.06  \\
  FastSpeech 2~\cite{ren2020fastspeech}           & 3.92$\pm$0.05   & 4.06  & 0.23 & 0.99 & 0.79 & 0.021  &0.01  \\
  GANSpeech~\cite{yang2021ganspeech}              & 4.00$\pm$0.05   & 4.02  & 0.21 & 0.96 & 0.73 & 0.104  &0.02  \\
  Glow-TTS~\cite{kim2020glow}                     & 4.01$\pm$0.07   & 4.35  & 0.19 & 1.00 & 0.74 & 0.012  &0.01 \\
  Grad-TTS (64 steps)~\cite{popov2021grad}        & 4.05$\pm$0.06   & 3.36  & 0.19 & 1.48 &\bfseries 0.57 & 0.023  &0.19\\
  DiffSpeech (128 steps)~\cite{liu2021diffsinger}  & \bfseries 4.09$\pm$0.06 & 3.48  & 0.83 & 2.40 & 0.67 & \bfseries 0.008  &1.11 \\
  \midrule
  \bfseries ProDiff (2 steps) & \bfseries 4.08$\pm$0.07 & \bfseries 3.15 &  \bfseries 0.85 & \bfseries 2.55 & 0.69 & 0.012  & \bfseries 0.04\\     
  \bottomrule
  \end{tabular}
  \vspace{1mm}
  \caption{The audio quality, sampling speed and sample diversity comparisons. The evaluation is conducted on a server with 1 NVIDIA 2080Ti GPU and batch size 1. For sampling, we use 64 steps in Grad-TTS and 128 steps in DiffSpeech, respectively, following~\cite{github2021Speech-Backbones} and ~\cite{github2021DiffSinger}.}
 
  \label{table:mos2}
  \end{table*}

\subsection{Performances} 
We compare the quality of generated audio samples, inference latency, and sample diversity with other systems, including 1) GT, the ground-truth audio; 2) GT (voc.), where we first convert the ground-truth audio into mel-spectrograms and then convert them back to audio using HiFi-GAN (V1)~\cite{kong2020hifi}; 3) Tacotron 2~\cite{shen2018natural}: the traditional autoregressive TTS model; 4) FastSpeech 2~\cite{ren2020fastspeech}: one of the most popular non-autoregressive TTS models; 5) Glow-TTS~\cite{kim2020glow}: the TTS model with the normalizing flow and monotonic alignment search; 6) GANSpeech~\cite{yang2021ganspeech}: the TTS model with generative adversarial networks; 7) Grad-TTS~\cite{popov2021grad} and DiffSpeech~\cite{popov2021grad}: two denoising diffusion probabilistic models. The results are compiled and presented in Table~\ref{table:mos2}, and we have the following observations: 

\begin{figure*}[ht]
    \centering
    \small
    \includegraphics[width=.9\textwidth]{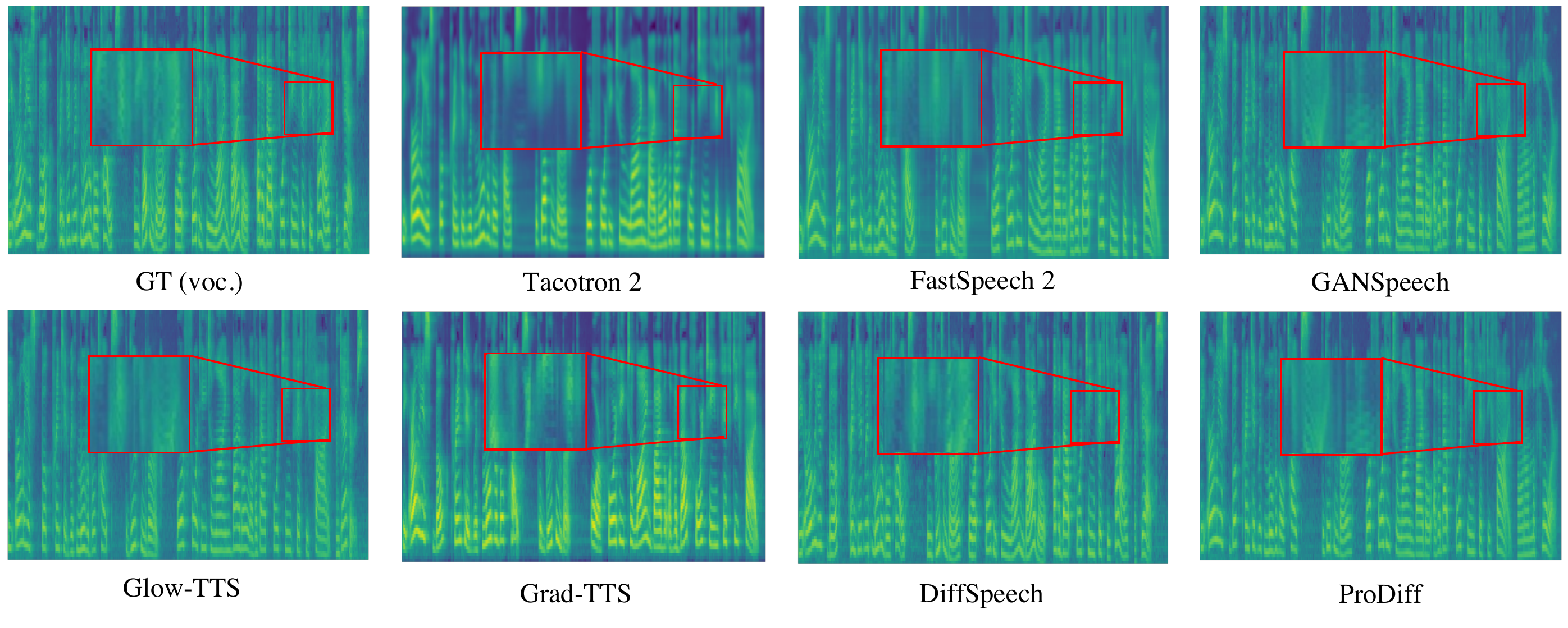}
    \caption{Visualizations of the ground-truth and generated mel-spectrograms by different TTS models. The corresponding text is "the earliest book printed with movable types, the gutenberg, or forty two line bible of about fourteen fifty five".}
    \label{fig:vis_mel}
    \end{figure*}

\textbf{Audio Quality}
In terms of audio quality, ProDiff achieves high perceptual quality with a gap of $0.17$ compared to the ground truth audio. It matches the SOTA DDPMs using hundreds of steps and outperforms other non-autoregressive baselines. For objective evaluation, ProDiff also demonstrates the outperformed performance in MCD, PESQ, and STOI, superior to all baseline models.

\textbf{Sampling Speed}
ProDiff enjoys an effective sampling process that needs only \textbf{2} iterations to synthesize high-fidelity spectrograms, 24x faster than real-time on a single NVIDIA 2080Ti GPU. ProDiff significantly reduces the inference time compared with the competing diffusion models (Grad-TTS and DiffSpeech). 

\begin{figure*}[ht]
    \centering
    \small
    \hspace{-5mm}
    \subfigure[Inference time.]
    {
        \label{fig:speed}
    \includegraphics[width=.4\textwidth]{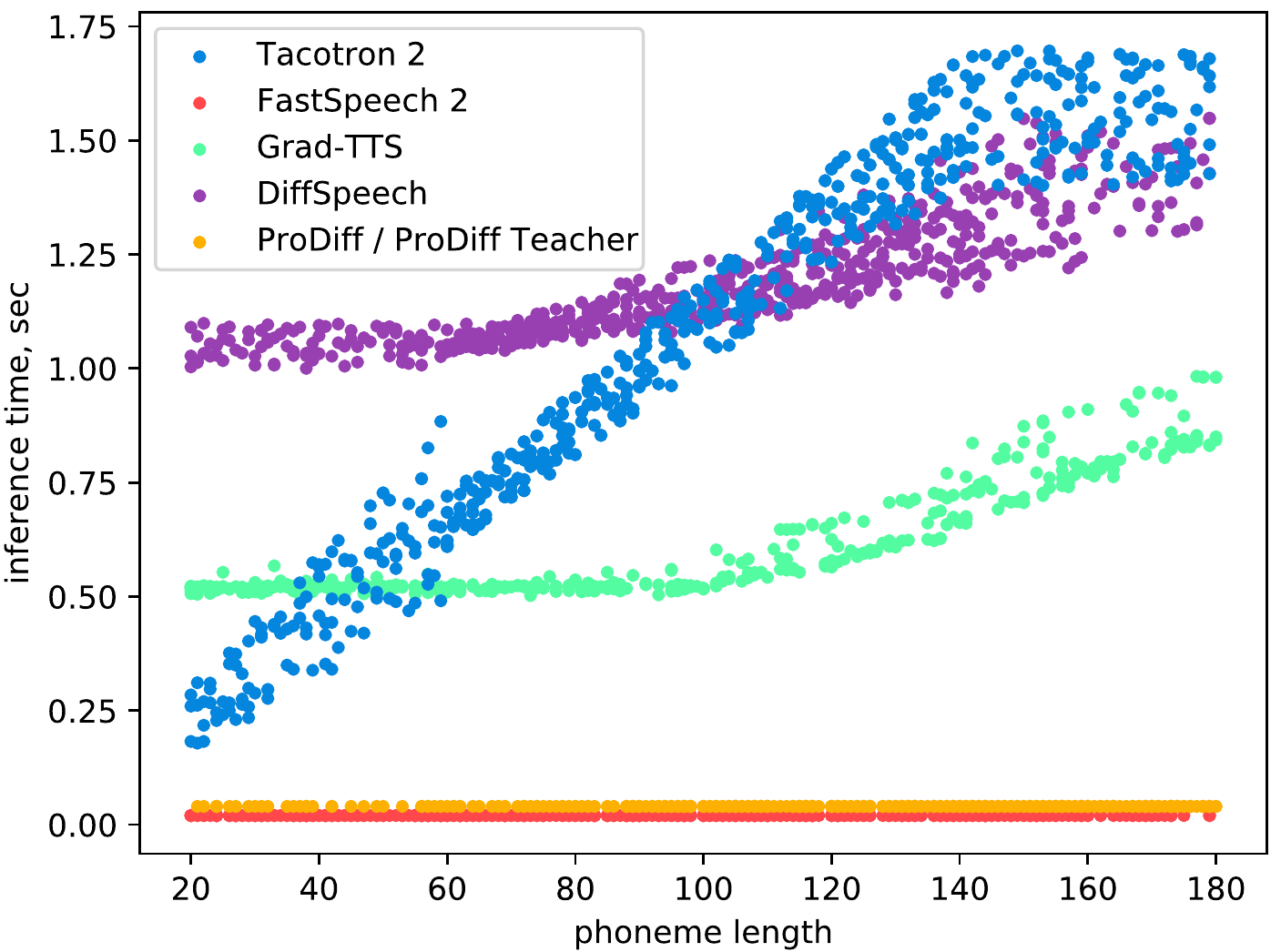}
    }
    \hspace{5mm}
    \subfigure[MCD evaluation results.]
    {
        \label{fig:mcd}
    \includegraphics[width=.4\textwidth]{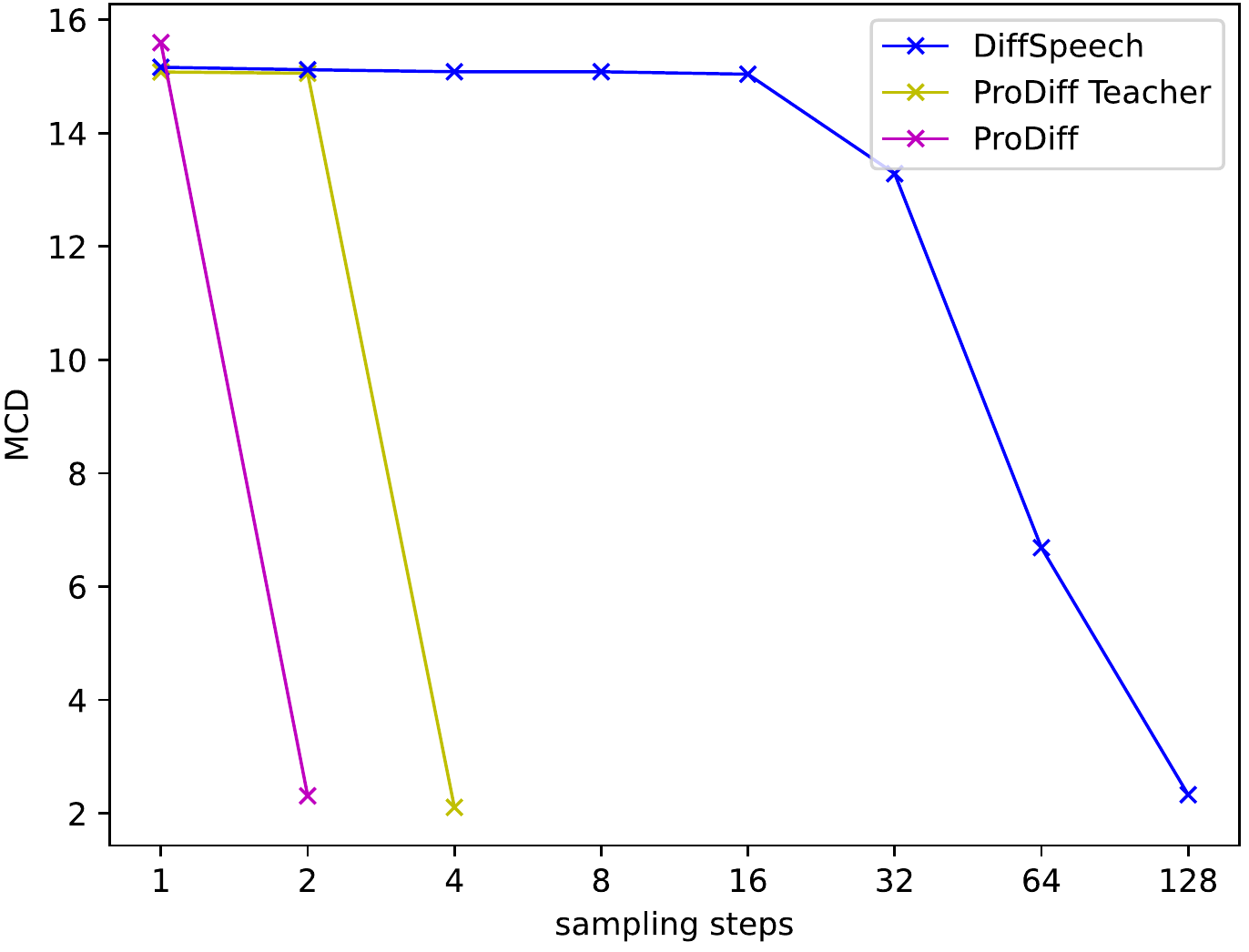}
    }
\caption{In subfigure (a), we evaluate the relationship between the inference latency and the length of the predicted mel-spectrogram sequence in the test set. In subfigure (b), we report MCD results obtained after each iteration of the reverse denoising algorithm.}
\end{figure*}

We visualize the relationship between the inference latency and the length of phoneme sequence. Figure~\ref{fig:speed} shows that the inference latency of ProDiff barely increases with the length of input phoneme, which is almost constant at 20ms. Unlike Tacotron 2, Grad-TTS, and DiffSpeech which linearly increase with the length due to autoregressive sampling or a large number of iterations, ProDiff has a similar scaling performance to FastSpeech 2, making diffusion models practically applicable to text-to-speech deployment for the first time.

\textbf{Sample Diversity}
Previous diffusion models~\cite{dhariwal2021diffusion,xiao2021tackling} in image generation have demonstrated the outperformed sample diversity, while the comparison in the speech domain is relatively overlooked. Similarly, we can intuitively infer that diffusion probabilistic models are good at generating diverse samples. To verify our hypothesis, we employ two metrics NDB and JS~\cite{richardson2018gans} to explore the diversity of generated mel-spectrograms. As shown in Table~\ref{table:mos2}, we can see that ProDiff achieves higher scores than several one-shot methods, while the GAN-based method generates samples with minimal diversity, which is expected for the following reasons:
 
1) It is well-known that mode collapse~\cite{creswell2018generative} easily appears in the strongly conditional generation task in one-shot generative models, leading to very similar output samples from a single or few modes of the distribution.

2) In contrast, diffusion models (Grad-TTS, DiffSpeech, and ProDiff) are meant to reduce mode collapse. They break the generation process into several conditional diffusion steps where each step is relatively simple to model. Thus, we expect our model to exhibit better training stability and mode coverage.

Besides, we conduct a quality comparison on the multi-speaker dataset and obtain similar conclusions as above. (see Appendix~\ref{appendix:multispeaker}). We also conduct robustness evaluation on both single-speaker and multi-speaker datasets in Appendix~\ref{appendix:robust} and find that ProDiff achieves comparable robustness performance with state-of-the-art non-autoregressive TTS models.

\subsection{Visualizations}
As illustrated in Figure~\ref{fig:vis_mel}, we then visualize the mel-spectrograms generated by the above systems given the same text sequence and have the following observations: Non-probabilistic models (Tacotron 2, FastSpeech 2) tend to generate less-diverse samples of blurry and over-smooth mel-spectrograms, and the GAN-based model (GANSpeech) suffers from model collapse with very similar output samples. Differently, diffusion probabilistic models (Grad-TTS, DiffSpeech, ProDiff) generate mel-spectrograms with rich frequency details, resulting in the natural and expressive sounds. ProDiff is demonstrated to make a sharp prediction with knowledge distillation, and it maintains sample quality competitive with state-of-the-art models using hundreds of steps.
   
\begin{table}[ht]
    \centering
    \begin{tabular}{lcccc}
    \toprule
    \bfseries Method  &\bfseries MOS ($\uparrow$)&\bfseries MCD ($\downarrow$) &\bfseries STOI ($\uparrow$)&\bfseries PESQ ($\uparrow$)\\
    \midrule
    GT(voc.)                         &4.25$\pm$0.06  & 1.08  & 0.95 & 3.18 \\
    \midrule
    ProDiff                         &  \bfseries 4.08$\pm$0.07 &  \bfseries 3.15 &  \bfseries 0.85 & \bfseries  2.55  \\
    w/o GP                          & 2.96$\pm$0.08 & 6.78 & 0.40 & 0.68 \\
    w/o KD                          & 3.92$\pm$0.04 & 3.23 & 0.81 & 2.51 \\
    \midrule
    Teacher (T=16)                   & 4.05$\pm$0.05 & 3.33 & 0.82 & 2.32 \\
    Teacher (T=8)                    & 4.06$\pm$0.06 & 3.18 & 0.85 & 2.50 \\
    \bottomrule
    \end{tabular}
    \vspace{2mm}
    \caption{Ablation study results. Comparison of the effect of each component in terms of quality and synthesis speed. We use GP to denote generator parameterization, and KD to denote knowledge distillation.}
    \label{table:mos3}
    \end{table}
     
\subsection{Progressive Diffusion}
To evaluate the efficiency of diffusion models in reverse sampling, we report MCD results obtained after each iteration. As shown in Figure~\ref{fig:mcd}, we compare against the highly optimized stochastic baseline sampler DiffSpeech: 1) DiffSpeech slowly refines the coarse-grained mel-spectrogram from Gaussian noise, and a guarantee of high sample quality comes at the cost of hundreds of denoising steps; 2) In contrast, ProDiff Teacher and ProDiff predict unperturbed $\vx_{0}$ and then add back perturbation using the posterior distribution, which produces near-optimal results more faster and effective, with attractive solutions for computational budgets that allow fewer reverse iterations.

\subsection{Ablation Studies}
We conduct ablation studies to demonstrate the effectiveness of several key techniques in ProDiff, including the generator-based diffusion parameterization and knowledge distillation. The results of both subjective and objective evaluations have been presented in Table~\ref{table:mos3}, and we have the following observations: 1) With limited diffusion iterations, replacing the generator-based diffusion parameterization by gradient-based parameterization causes a distinct degradation in perceptual quality. ProDiff directly predicts clean data to avoid significant degradation when reducing reverse iterations. 2) Removing the knowledge distillation mechanism and using clean data as training target results in blurry and over-smooth predictions, which demonstrates the effectiveness and efficiency of the proposed distillation in reducing data variance and promoting model convergence.

Furthermore, we compare to distill the behavior of a teacher with varying diffusion steps (i.e., 16, 8, and 4) into a 2-step student. Take the 16-step teacher as an example, we need 3 distillation procedures to obtain the student (16->8->4->2), where the student model iteratively becomes the teacher in the following distillation. However, the perceptual quality of generated samples slightly drops after a series of distillations, indicating that the quality gain from sharper prediction could hardly cover the degradation in reducing iterations. In summary, distilling knowledge from the 4-step teacher could be an optimal choice to strike a balance between computational cost and sample quality.

\section{Conclusion}
In this work, we proposed ProDiff, on progressive fast diffusion model for high-quality text-to-speech. The preliminary study on diffusion model parameterization found that previous gradient-based TTS models required hundreds of iterations to guarantee high sample quality, which posed a challenge for accelerating sampling. In contrast, ProDiff parameterized the denoising model by directly predicting the clean data to avoid significant degradation of perceptual quality when reducing reverse iterations. To tackle the model convergence challenge in accelerating refinement, ProDiff adopted the synthesized mel-spectrogram from teacher as target to reduce the data variance and make a sharp prediction. As such, ProDiff was distilled from the behavior of an N-step teacher into a new model with N/2 steps, further reducing the sampling time by orders of magnitude.

Experimental results demonstrated that ProDiff needed only 2 iterations to synthesize high-fidelity mel-spectrograms, while it maintained sample quality and diversity competitive with state-of-the-art models using hundreds of steps. To the best of our knowledge, ProDiff made diffusion models for the first time applicable to interactive, real-world text-to-speech with a low computational cost. Our extensive ablation studies demonstrated that each design in ProDiff was effective, and we showed that our model could be easily extended to a multi-speaker setting. We envisage that our work could serve as a basis for future text-to-speech studies.

\section*{Acknowledgements}
The author would like to thank Luping Liu and Max W.Y. Lam for the helpful discussions over the initial ideas. This work was supported in part by the Zhejiang Natural Science Foundation LR19F020006 and National Key R\&D Program of China under Grant No.2020YFC0832505.

\bibliographystyle{ACM-Reference-Format}
\bibliography{sample-base}

\appendix

\section{Diffusion Posterior Distribution} \label{Posterior}
Firstly we compute the corresponding constants respective to diffusion and reverse process:
\begin{align}
    \alpha_{t}=\prod_{i=1}^{t} \sqrt{1-\beta_{i}} \quad \sigma_t=\sqrt{1-\alpha_t^2}
\end{align}

The Gaussian posterior in diffusion process is defined through the Markov chain, where each iteration adds Gaussian noise. Consider the forward diffusion process in Eq.~\ref{diffusion}, which we repeat here:
\begin{align}
    \begin{split}
q(\vx_{1},\cdots,\vx_T|x_0) &= \prod_{t=1}^T q(\vx_t|\vx_{t-1}),\\ q(\vx_t|\vx_{t-1})=& \gN(\vx_t;\sqrt{1-\beta_t}\vx_{t-1},\beta_t \rmI)
\quad\ \ 
\end{split}
\end{align}

We emphasize the property observed by~\cite{ho2020denoising}, the diffusion process can be computed in a closed form:
\begin{equation}
    q(\vx_{t}|\vx_{0})=\gN(\vx_{t} ; \alpha_{t} \vx_{0},\sigma_t \rmI)
\end{equation}

Applying Bayes' rule, we can obtain the forward process posterior when conditioned on $\vx_0$
\begin{align}
    \begin{split}
    q(\vx_{t-1}|\vx_{t}, \vx_{0}) &=\frac{q(\vx_{t}|\vx_{t-1}, \vx_{0}) q(\vx_{t-1}|\vx_{0})}{q(\vx_{t}|\vx_{0})} \\
    =\frac{q(\vx_{t}|\vx_{t-1}) q(\vx_{t-1}|\vx_{0})}{q(\vx_{t}|\vx_{0})} & =\gN(\vx_{t-1};\tilde{\boldsymbol{\mu}}_{t}(\vx_{t}, \vx_{0}), \tilde{\beta}_{t} I),
\end{split}
\end{align}
where $\tilde{\boldsymbol{\mu}}_{t}(\vx_{t}, \vx_{0})=\frac{\alpha_{t-1} \beta_{t}}{\sigma_t} \vx_{0}+\frac{\sqrt{1-\beta_{t}}(\sigma_{t-1})}{\sigma_t} \vx_{t},  \quad \tilde{\beta}_{t}=\frac{\sigma_{t-1}}{\sigma_t} \beta_{t}$

\section{Distill Target}
Inspired by~\citep{salimans2022progressive}, to promote the perceptual quality of generated sample, ProDiff reduces the data variance in the target side via knowledge distillation. The denoising model use the generated mel-spectrogram from an N-step DDIM teacher as the training target, and distill the behavior into a new model with N/2 steps. 

In this section, the student model needs to predict the specific target in order to match the teacher during sampling. Here we derive what this target needs to be:
When training the N-step student, we have that the teacher model samples the next set of noisy data $\vx_{t^{\prime \prime}}$ given the current noisy data $\vx_{t}$ by taking two steps of DDIM. The student tries to sample the same value in only one step of DDIM. Denoting the student denoising prediction by $\hat \vx_0$, and its one-step sample by $\hat \vx_t$, we gives:

\begin{equation}
    \tilde{\vx}_{t^{\prime \prime}}=\alpha_{t^{\prime \prime}} \hat \vx_0 +\frac{\sigma_{t^{\prime \prime}}}{\sigma_{t}}(\vx_{t}-\alpha_{t} \hat \vx_0)
    \end{equation}

In order for the student sampler to match the teacher sampler, we must set $\tilde{\vx}_{t^{\prime \prime}}$ equal to $\vx_{t^{\prime \prime}}$ . This gives

\begin{align}
\begin{split}
    \tilde{\vx}_{t^{\prime \prime}} &=\alpha_{t^{\prime \prime}} \hat \vx_0+\frac{\sigma_{t^{\prime \prime}}}{\sigma_{t}}\left(\vx_{t}-\alpha_{t} \hat \vx_0\right)=\vx_{t^{\prime \prime}}\\
        &=\left(\alpha_{t^{\prime \prime}}-\frac{\sigma_{t^{\prime \prime}}}{\sigma_{t}} \alpha_{t}\right) \hat \vx_0+\frac{\sigma_{t^{\prime \prime}}}{\sigma_{t}} \vx_{t}=\vx_{t^{\prime \prime}} \\
        \left(\alpha_{t^{\prime \prime}}-\frac{\sigma_{t^{\prime \prime}}}{\sigma_{t}} \alpha_{t}\right) \hat \vx_0 &=\vx_{t^{\prime \prime}}-\frac{\sigma_{t^{\prime \prime}}}{\sigma_{t}} \vx_{t} \\
        \hat \vx_0 &=\frac{\vx_{t^{\prime \prime}}-\frac{\sigma_{t^{\prime \prime}}}{\sigma_{t}} \vx_{t}}{\alpha_{t^{\prime \prime}}-\frac{\sigma_{t^{\prime \prime}}}{\sigma_{t}} \alpha_{t}}
\end{split}
\end{align}

\section{Architecture} \label{appendix:arch}
We list the model hyper-parameters of ProDiff in Table~\ref{tab:hyperparameters_ps}.

\begin{table}[h]
\small
\centering
\begin{tabular}{l|c|c}
\toprule
\multicolumn{2}{c|}{Hyperparameter}   & ProDiff \\ 
\midrule
\multirow{9}{*}{Text Encoder} 
&Phoneme Embedding           &256   \\
&Pre-net Layers              &3   \\
&Pre-net Hidden              &256   \\
&Encoder Layers              &4   \\
&Encoder Hidden              &256     \\                      
&Encoder Conv1d Kernel       &9   \\    
&Encoder Conv1D Filter Size  &1024\\                 
&Encoder Attention Heads     &2   \\    
&Encoder Dropout             &0.05\\                       
\midrule
\multirow{3}{*}{Variance Predictor}         
&Variance Predictor Conv1D Kernel        & 3\\    
&Variance Predictor Conv1D Filter Size   & 256  \\    
&Variance Predictor Dropout              & 0.5  \\ 
\midrule
\multirow{5}{*}{Spectrogram Denoiser}     
&Diffusion Embedding                &  256  \\   
&Residual Layers                    &  20 \\       
&Residual Channels                  &  256 \\     
&WaveNet Conv1d Kernel              &  3 \\       
&WaveNet Conv1d Filter              &  512 \\ 
\midrule
\multicolumn{2}{c|}{Total Number of Parameters}   & 17M  \\
\bottomrule
\end{tabular}
\caption{Hyperparameters of ProDiff models.}
\label{tab:hyperparameters_ps}
\end{table}

\section{Training algorithm}

\begin{algorithm}[ht]
    \centering
    \caption{Training gradient-based diffusion model}\label{alg: training_gradient}
    \begin{algorithmic}[1]
     \STATE \textbf{Require}: Diffusion model $\theta$.
    \REPEAT 
    \STATE Sample $\vx_{0} \sim q_{data}$, $\rvepsilon\sim\gN(\vzero,\mI)$, and $t\sim\mathrm{Unif}(\{0,\cdots,T\})$
    \STATE $\vx_t = \alpha_{t} \vx_{0}+ \sigma_t \rvepsilon$
    \STATE Take gradient descent steps on $\nabla_{\theta}(\gL_{\theta}^{\text{Grad}} + \gL_{\operatorname{SSIM}} + \gL_{p} + \gL_{e} + \gL_{dur})$.
    \UNTIL{Diffusion model $\theta$ converged}
    \end{algorithmic}
    \end{algorithm}

\begin{algorithm}[ht]
    \centering
    \caption{Training generator-based diffusion model}\label{alg: training_generator}
    \begin{algorithmic}[1]
     \STATE \textbf{Require}: Diffusion model $\theta$.
    \REPEAT 
    \STATE Sample $\vx_{0} \sim q_{data}$, $\rvepsilon\sim\gN(\vzero,\mI)$, and $t\sim\mathrm{Unif}(\{0,\cdots,T\})$
    \STATE $\vx_t = \alpha_{t} \vx_{0}+ \sigma_t \rvepsilon$
    \STATE Take gradient descent steps on $\nabla_{\theta}(\gL_{\theta}^{\text{Gen}} + \gL_{\operatorname{SSIM}} + \gL_{p} + \gL_{e} + \gL_{dur})$.
    \UNTIL{Diffusion model $\theta$ converged}
    \end{algorithmic}
    \end{algorithm}

\section{Results on Multi-Speaker Dataset}  \label{appendix:multispeaker}
For the multi-speaker setting, the train-clean-100 subset of the LibriTTS corpus~\cite{zen2019libritts} is used, which consists of audio recordings of 247 speakers with a total duration of about 54 hours. We add speaker embedding after the encoder and before the spectrogram denoiser, following the common practice~\cite{kim2020glow,min2021meta}. The results have been compiled in Table~\ref{table:mos4}. Similar to the experimental results on LJSpeech, we can draw conclusions that ProDiff achieve outperformed quality in terms of both subjective and objective evaluation, even in more complicated (multi-speaker) scenarios.

\begin{table}[ht]
    \begin{tabular}{lcccc}
    \toprule
    \bfseries Method & \bfseries MOS ($\uparrow$) & \bfseries MCD $\downarrow$) & \bfseries STOI ($\uparrow$) & \bfseries PESQ ($\uparrow$) \\
    \midrule
    GT              & 4.32$\pm$0.07 &  /      &   /     &    /       \\
    GT(voc.)        & 4.17$\pm$0.06  &  0.91      & 0.94	       &  3.26        \\
    \midrule
    Tacotron 2~\cite{shen2018natural}      & 3.82$\pm$0.07  &  5.45      &  0.09      &  0.79   \\
    FastSpeech 2~\cite{ren2020fastspeech}    & 3.87$\pm$0.06  & 4.67       &  0.11      & 0.81   \\
    GANSpeech~\cite{yang2021ganspeech}       & 3.95$\pm$0.06  & 4.86       &  0.17      & 1.11    \\
    Glow-TTS~\cite{kim2020glow}        & 3.97$\pm$0.05  &  4.84      &  0.16      &  0.97  \\
    Grad-TTS~\cite{popov2021grad}        & 4.00$\pm$0.07  & 5.25       & 0.12       &  0.92   \\
    DiffSpeech~\cite{popov2021grad}      & \textbf{4.06$\pm$0.06}  & 4.12       & 0.51       & 1.83    \\
    \midrule
    \bfseries ProDiff  & 4.04$\pm$0.06 &\textbf{4.01} & \textbf{0.54}& \textbf{1.97}    \\    
    \bottomrule
    \end{tabular}
    \caption{The audio performance comparisons in multi-speaker scenarios.}
    \label{table:mos4}
    \end{table}

\section{Robustness Evaluation} \label{appendix:robust}
We conduct the robustness evaluation on LJSpeech and LibriTTS datasets. We select 50 sentences that are particularly hard for TTS systems following FastSpeech~\cite{ren2019fastspeech}. The results are shown in Tables~\ref{table:robust1} and ~\ref{table:robust2}. We can see that ProDiff achieve comparable robustness performance with state-of-the-art non-autoregressive TTS models.

\begin{table}[]
    \begin{tabular}{lccc}
    \toprule
    \bfseries Method & \bfseries Repeats & \bfseries Skips & \bfseries Error Sentences \\
    \midrule
    Tacotron 2~\cite{shen2018natural}    & 5         & 4      & 7                \\
        FastSpeech 2~\cite{ren2020fastspeech}& 1        & 0      & 1                \\
        GANSpeech~\cite{yang2021ganspeech}   & 1        & 2      & 2                \\
        Glow-TTS~\cite{kim2020glow}         & 1        & 1   & 2                \\
        Grad-TTS~\cite{popov2021grad}        & 0        & 2      & 2                \\
        DiffSpeech~\cite{popov2021grad}      & 1        & 1      & 1                \\
        \midrule
        \bfseries ProDiff                       & 2       & 1       & 2     \\
    \bottomrule
    \end{tabular}
    \caption{The robustness evaluation on LJSpeech dataset.}
    \label{table:robust1}
    \end{table}

\begin{table}[]
    \begin{tabular}{lccc}
        \toprule
        \bfseries Method & \bfseries Repeats & \bfseries Skips & \bfseries Error Sentences \\
        \midrule
        Tacotron 2~\cite{shen2018natural}    & 7         & 6      & 12                \\
        FastSpeech 2~\cite{ren2020fastspeech}& 1        & 2      & 2                \\
        GANSpeech~\cite{yang2021ganspeech}   & 2        & 2      & 3                \\
        Glow-TTS~\cite{kim2020glow}         & 4        & 5   & 8                \\
        Grad-TTS~\cite{popov2021grad}        & 3        & 5      & 6                \\
        DiffSpeech~\cite{popov2021grad}      & 1        & 3      & 3                \\
        \midrule
        \bfseries ProDiff                       & 2       & 3       & 4     \\
        \bottomrule
        \end{tabular}
    \caption{The robustness evaluation on LibriTTS dataset.}
    \label{table:robust2}
    \end{table}

\section{Evaluation Matrix} \label{appendix:evaluation}

\subsection{Subjective Evaluation}
All our Mean Opinion Score (MOS) tests are crowdsourced and conducted by native speakers. We refer to the rubric for MOS scores in~\cite{protasio_ribeiro_crowdmos_2011}, and the scoring criteria has been included in Table~\ref{matrix:naturalness} for completeness. The samples are presented and rated one at a time by the testers. 

\begin{table}[H]
  \begin{tabular}{ccc}
  \toprule
  Rating & Naturalness & Definition                           \\
  \midrule
  1      & Bad        &  Very annoying and objectionable dist. \\
  2      & Poor       &  Annoying but not objectionable dist. \\
  3      & Fair       &  Perceptible and slightly annoying dist\\
  4      & Good       & Just perceptible but not annoying dist. \\
  5      & Excellent  & Imperceptible distortions\\
  \bottomrule
  \end{tabular}
  \caption{Ratings that have been used in evaluation of speech naturalness of synthetic and ground truth samples.}
    \label{matrix:naturalness}
\end{table}

\begin{figure*}[ht]
    \vspace{-3mm}
    \centering
    \subfigure[Gradient-based diffusion parameterization]
    {
    \label{fig:Generator}
    \includegraphics[width=.9\textwidth]{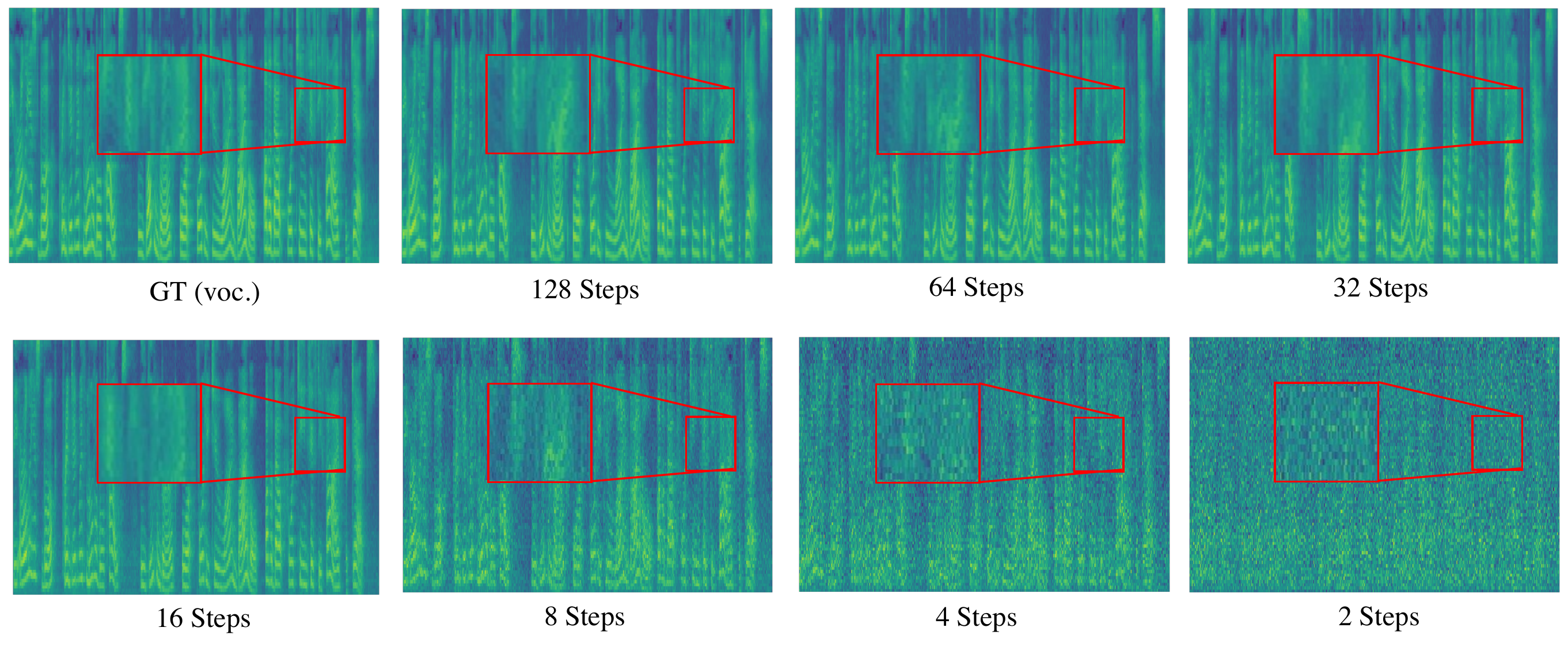}
    }
    \subfigure[Generator-based diffusion parameterization]
    {
    \label{fig:Discriminator}
    \includegraphics[width=.9\textwidth]{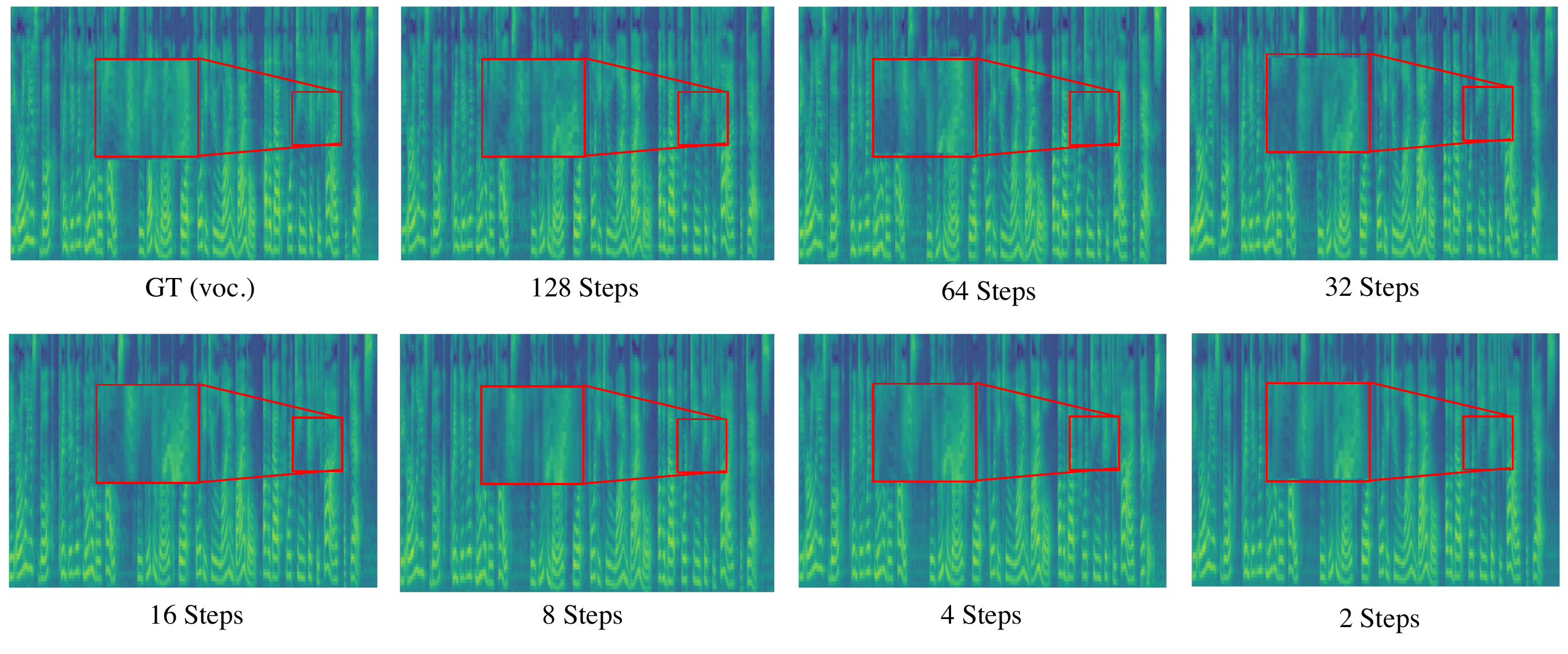}
    }
     \vspace{-3mm}
\caption{Visualizations of the ground-truth and generated mel-spectrograms by different TTS models with varying diffusion steps. The corresponding text is "the earliest book printed with movable types, the gutenberg, or forty two line bible of about fourteen fifty five".}
\label{fig:vis_parameterization} 
\end{figure*}

\subsection{Objective Evaluation} 
Mel-cepstral distortion (MCD)~\cite{kubichek1993mel} measures the spectral distance between the synthesized and reference mel-spectrum features.

Perceptual evaluation of speech quality (PESQ)~\cite{rix2001perceptual} and The short-time objective intelligibility (STOI)~\cite{taal2010short} assesses the denoising quality for speech enhancement.

Number of Statistically-Different Bins (NDB) and Jensen-Shannon divergence (JSD)~\cite{richardson2018gans}. They measure diversity by 1) clustering the training data into several clusters, and 2) measuring how well the generated samples fit into those clusters.

\section{Visualizations} \label{appendix:visualizations}

\end{document}